\documentstyle[sprocl]{article}

\input{psfig}
\def\tr{{\rm tr}}
\def\A{{\bf A}}
\def\B{{\bf B}}
\def\D{{\bf D}}
\def\E{{\bf E}}
\def\J{{\bf J}}
\def\x{{\bf x}}
\def\grad{\mbox{\boldmath$\nabla$}}
\def\large{{\rm large}}
\def\any{{\rm any}}

\def\paper{1}
\def\ASYtwonum{5}

\makeatletter

\def\gtrsim{\mathrel{\mathpalette\vereq>}}
\def\vereq#1#2{\lower3pt\vbox{\baselineskip1.5pt \lineskip1.5pt
\ialign{$\m@th#1\hfill##\hfil$\crcr#2\crcr\sim\crcr}}}
\makeatother

\begin{document}

\title{RATES FOR NON-PERTURBATIVE PROCESSES IN HOT NON-ABELIAN PLASMAS%
\footnote{
   Talk given at ``Continuous Advances in QCD '98.''}}

\author{PETER ARNOLD}

\address{
    Department of Physics,
    University of Virginia,
    Charlottesville, VA 22901 \\
    ~\\
{\rm\normalsize overviewing work done\cite{paper1} in collaboration with}
}

\author{DAM SON}
\address
    {%
    Center for Theoretical Physics,
    Department of Physics, \\
    Massachusetts Institute of Technology,
    Cambridge, MA 02139
    }%
\author{LAURENCE G. YAFFE}
\address
    {%
    Department of Physics,
    University of Washington,
    Seattle, Washington 98195
    }%

\maketitle\abstracts{
   I give a simple physical picture of why
   the rate $\Gamma$ per unit volume of
   non-perturbative processes
   in hot non-Abelian plasmas (such as electroweak baryon number
   violation in the very early universe) depends on coupling as
   $\Gamma \sim \alpha^5 T^4 \ln(1/\alpha)$.  The logarithmic
   dependence was recently discovered by B\"odeker.
}

\section {Introduction}

   I will be discussing the rate for non-perturbative processes in hot
non-Abelian plasmas.  For me ``hot'' means high enough temperature that
the running coupling
$\alpha(T)$ is small, that chemical potentials are ignorable, and that
there is no spontaneous symmetry breaking.  Examples of non-perturbative
processes include chirality violation in hot QCD and baryon number
violation in hot electroweak theory (in its high-temperature symmetric
phase).  Both cases are related by the anomaly
\begin {equation}
   \Delta Q \sim g^2 \int d^4 x \> \tr F \tilde F
\end {equation}
to non-perturbative fluctuations of the gauge fields.  The question then
becomes: what is the rate for such non-perturbative fluctuations?

In fact, it turns out that a non-trivial question is how the rate
$\Gamma$ per unit volume even depends on the coupling $\alpha$
for small coupling:
\begin {equation}
   \Gamma = \# \alpha^{\hbox{\small{\bf ?}}} T^4
   ~~~~~~\hbox{(+ higher-order)} .
\end {equation}
The $T^4$ dependence just comes from dimensional analysis.  In this
talk, I shall review the spatial and temporal scales characteristic of
non-perturbative physics in the hot plasma, showing that the
distance scale is
\begin {equation}
   R \sim {1\over g^2 T}
\label{eq:Rscale}
\end {equation}
and the time scale is
\begin {equation}
   t \sim {1 \over g^4 T \ln(1/g)} .
\label{eq:tscale}
\end {equation}
If you accept these scales on faith for a moment, then one will have,
roughly,
one unsuppressed non-perturbative process per volume $R^3$ per time $t$,
so that
\begin {equation}
   \Gamma \sim {1 \over R^3 t} \sim \alpha^5 T^4 \ln\left(1\over\alpha\right)
   .
\label{eq:Gscale}
\end {equation}
The $\alpha^5$ is a result of my collaborators and I from two years
ago.\cite{ASY}
The logarithmic enhancement is a recent
and interesting result of Dietrich B\"odeker.\cite{Bodeker}

\section {The Spatial Scale}

Imagine a non-perturbative fluctuation of the gauge fields with some spatial
extent $R$.  Non-perturbative means that $gA$ is not a perturbation in
the covariant derivative $D = \partial - gA$ and so
\begin {equation}
   A \sim {1\over gR} .
\end {equation}
The energy of the configuration is then order
\begin {equation}
   E \sim {1\over g^2 R} \, .
\end {equation}
The probability of such a fluctuation at finite temperature $T$ is,
roughly speaking, given by a Maxwell-Boltzmann factor:
\begin {equation}
   e^{-\beta E} \sim e^{-1 / (g^2 R T)} ,
\end {equation}
and this probability is unsuppressed when
\begin {equation}
   R \gtrsim {1\over g^2 T} .
\end {equation}
In fact, because of entropy, the {\it smallest} unsuppressed distance scale
is the dominant one---there are more possible small configurations than
big ones.  So the characteristic scale of non-perturbative fluctuations is
$R \sim 1/g^2 T$
as claimed in the introduction, corresponding to a spatial momentum scale
of
\begin {equation}
   k \sim g^2 T .
\label {eq:kscale}
\end {equation}

Some readers may wonder how non-perturbative physics in the hot plasma is
consistent with deconfinement---something that's supposed to happen when you
heat up non-Abelian gauge theory.  In a hot plasma, static electric fields are
screened by the Debye effect, which arises from the response of the charges in
the plasma to the electric field.  This screening of the electric field is
what is meant when people speak of ``deconfinement''---it means that
the effective potential between static test charges is not confining.
However, static magnetic fields are {\it not} screened in a plasma (which is
why, for instance, the galaxy can have a magnetic field).  The physics
associated with static or nearly static non-Abelian magnetic fields can still
be non-perturbative.%
\footnote{
   On a more technical level, large space-like Wilson loops still have
   area-law behavior at high temperature.
}

\section{The time scale}

It was long assumed in the literature on electroweak baryogenesis that
the time scale $t$ was the same as the spatial scale $R$.  This need
not be true because the presence of the plasma breaks Lorentz invariance.
In fact, the dynamics is slowed down because plasmas are conductors
and because of
Lenz's Law: a conductive
medium resists changes in magnetic field.
This qualitative observation, in this context, is due to Guy
Moore.\cite{private}
Now to understand the time scale, start with Ampere's Law:
\begin {equation}
   \grad \times \B = {d \E\over dt} + \J = {d\E\over dt} + \sigma \E ,
\end {equation}
where $\sigma$ is the (color) conductivity of the plasma and where, for
the sake of notational and conceptual simplicity in these rough arguments,
I will simply write familiar Abelian
equations ({\it e.g.} $\grad\times\B$ instead of $\D\times\B$).
Let's anticipate the end result that the dynamics will be slow and
ignore the $d\E/dt$ term above.  In $A_0 = 0$ gauge, the result
may be written as
\begin {equation}
   \grad \times \grad \times \A \simeq - \sigma {d\A\over dt} .
\end {equation}
In terms of characteristic scales, this means
\begin {equation}
   {1\over R^2} A \sim \sigma {1\over t} A ,
\end {equation}
and so the characteristic time scale is
\begin {equation}
   t \sim \sigma R^2 .
\label {eq:t}
\end {equation}
As I shall review, the color conductivity of a non-Abelian plasma is
of order
\begin {equation}
   \sigma \sim {T \over \ln(1/\alpha)} .
\end {equation}
Combined with the spatial scale (\ref{eq:Rscale}), this leads to the results
(\ref{eq:tscale}) and (\ref{eq:Gscale}) claimed for the time scale
and the rate $\Gamma$ in the introduction.
A more proper and detailed discussion of the above and of what follows
may be found in ref.~\ASYtwonum.

\begin{figure}[t]
\begin {center}
\leavevmode
\psfig{figure=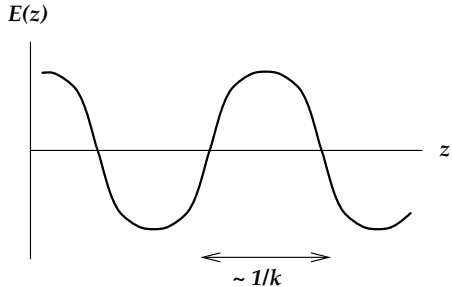,height=1.5in}
\end {center}
\caption{
  A spatially oscillating electric field with wave number $k$.
  \label{fig:Efield}}
\end{figure}

\section {The color conductivity}

\subsection{Collisionless Plasmas}

I'll now explain how one understands the conductivity in hot plasmas.
I'll begin by first considering the case of a collisionless plasma---that
is, what is the current response to an external electric field if we
ignore interactions between the charge carriers in the plasma?
Consider this response as a function of the wave number $k$.  Specifically,
imagine applying an external electric field of the form
$\E(\x) = \E_0 \cos(k z)$, such as depicted in fig.~\ref{fig:Efield}.
Charges in the plasma will be accelerated by the electric field,
creating a current.  If the electric field were spatially homogeneous,
the charge carriers would be accelerated indefinitely, and the current
would grow arbitrarily large in the absence of collision.  In the
case of the spatially oscillating electric field, however, this growth
of the current is cut off when the particles fly into a region where the
direction of $\E$ is reversed.  The larger $k$, the sooner this happens,
and the smaller the current.
The conductivity $\sigma = J/E$ turns out to be of order
\begin {equation}
   \sigma \sim {g^2 T^2 \over k} ,
\label{eq:collisionless}
\end {equation}
where the $1/k$ embodies the cut-off effect just described.  One factor
of $g$ is from the coupling of the charges to the electric field, the other
factor of $g$ is from the definition of the current as charge times
velocity.

If one plugs the spatial scale (\ref{eq:kscale}) of soft dynamics into
(\ref{eq:collisionless}), the order of the conductivity becomes
$\sigma \sim T$, which would give a time scale (\ref{eq:t}) of
$t\sim 1/g^4 T$ and rate $\Gamma \sim \alpha^5 T^4$.  This answer
misses B\"odeker's logarithmic enhancement.

\subsection {What about $k \to 0$?}

\begin{figure}[t]
\begin{center}
\leavevmode
\psfig{figure=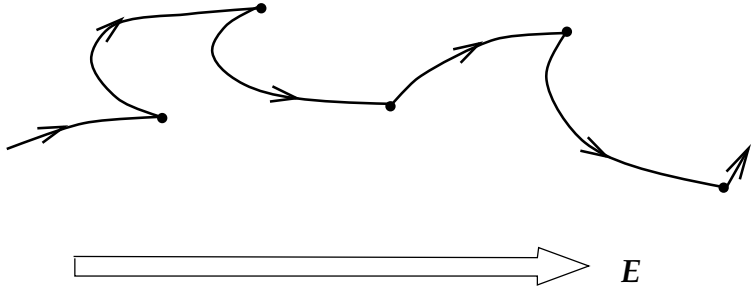,height=1.5in}
\end{center}
\caption{
  The path of a drunk heading towards a warehouse of wine.
  \label{fig:drunk}}
\end{figure}

First consider the case of an Abelian plasma.
The collisionless result (\ref{eq:collisionless}) obviously makes no
sense as we take $k \to 0$.  Real conductors, like copper, obviously
have a finite response to homogeneous electric fields.  The reason is
collisions.  Every so often, electrons have hard, large-angle
collisions that randomize their direction.  Each such randomization
of direction randomizes that electron's contribution to the current.
This situation is depicted in fig.~\ref{fig:drunk},
which shows the path taken
by a drunk who is being accelerated to the right by the smell of a
warehouse full of wine.  Every so often the drunk smashes into a
lamp post and stumbles away in a random direction.  As a result, the
drunk's velocity towards the warehouse is limited.

\begin{figure}[t]
\begin {center}
\leavevmode
\psfig{figure=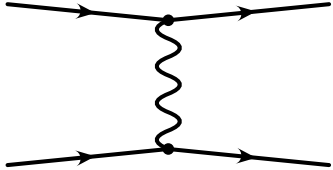,height=0.85in}
\end {center}
\caption{
  A collision between charge carriers.
  \label{fig:tchannel}}
\end{figure}

Here, the time scale that determines the limiting current is the
mean free time for the direction to be randomized.  In the collisionless
case, it was instead the time scale of order $1/k$ for particles to travel
from a region of $+\E_0$ to $-\E_0$.  A diagram for a collision is
shown in fig.~\ref{fig:tchannel}.  The square matrix element $|{\cal M}|^2$ is
explicitly of order $g^4$.  The mean free time for hard collisions is
then order
\begin {equation}
   \tau_\large \sim {1\over g^4 T \ln},
\end {equation}
where the $T$ is by dimensional analysis and the logarithm
turns out to arise because randomization of
direction can occur either through a single large-angle scattering or a
succession of (individually more probable) small-angle ones.
If this were the end of the story, then collisions would be irrelevant
for non-perturbative dynamics in non-Abelian plasmas, because the
collision time scale $\tau_\large$ is large compared to the time scale
$1/k \sim 1/g^2 T$ relevant to the previous section.  That is,
collisions would be rare enough that we could ignore them, at the
spatial scale of interest, and use the collisionless result
(\ref{eq:collisionless}) for the conductivity.

However, as first noted by Selikhov and Gyulassy,\cite{SG}%
\footnote{
   I am grateful to H. Heiselberg for pointing me towards the literature
   on this subject after my talk at the conference.
}
the situation is quite different for non-Abelian plasmas.
The reason is that the virtual gluon exchanged in fig.~\ref{fig:tchannel}
carries
color and so changes the colors of the colliding particles.
That means that very soft, small-angle collisions, even though they
do not individually randomize the {\it direction} of the charge
carriers, can instead randomize the {\it charge} of the charge
carriers.  Randomizing the charges, however, is enough to randomize
the individual currents.  Imagine that the drunk in fig.\ \ref{fig:drunk}
now has a multiple personality disorder: sometimes he's a drunk (``wine!  more wine!''), sometimes
he's a teetotaler (``Goodness Gracious, I must head home''), and he
vacillates between these personalities much faster than he has hard
collisions with lamp posts.%
\footnote{
   The author is fully aware that this analogy with the drunk serves no
   useful purpose---he just enjoys it.
}

The mean free time that determines the
color conductivity is therefore the mean free time for {\it any}
angle scatterings, not just large-angle ones.  The total cross-section
for any scattering suffers from the usual Coulomb divergence and
would formally be infinite if not for
screening effects in the plasma and non-perturbative effects.
When one includes such screening
effects, one finds that the mean free time is
\begin {equation}
   \tau_\any \sim {1\over g^2 T \ln} .
\end {equation}
This time is {\it smaller} than $1/k$ by a logarithm and so dominates
the physics when $\alpha$ is small enough that the logarithm is big.
Replacing the time scale $1/k$ by $\tau_\any$ in eq.\ (\ref{eq:collisionless})
for the conductivity, one obtains
\begin {equation}
   \sigma \sim {T \over \ln(1/\alpha)}
\end {equation}
as claimed earlier.  The fact that $\tau_\any$ beats $1/k$ by a logarithm
is what, in this language, is responsible for the logarithmic enhancement of
the rate $\Gamma$ in (\ref{eq:Gscale}), found by B\"odeker.

For a much more in depth discussion of the topics I have discussed, see
ref.~\paper.


\section*{References}
\begin{thebibliography}{99}

\bibitem{paper1}
  P. Arnold, D. Son, and L. Yaffe,
  {\tt hep-ph/9810216},
  Univ.\ of Washington report no.\ UW/PT 98--10.

\bibitem{ASY}
  P. Arnold, D. Son, and L. Yaffe,
  {\tt hep-ph/9609481},
  {\em Phys.\ Rev.}\ {\bf D55}, 6264 (1997).

\bibitem{Bodeker}
    D. B\"odeker,
    {\tt hep-ph/9810430},
    {\em Phys.\ Lett.}\ {\bf B426}, 351 (1998).

\bibitem{private}
   Guy Moore, private communication.

\bibitem{ASY2}
   P. Arnold, D. Son, and L. Yaffe,
   University of Washington preprint UW/PT 98-10.

\bibitem{SG}
    A. Selikhov and M. Gyulassy,
    {\tt nucl-ph/9307007},
    {\em Phys.\ Lett.}\ {\bf B316}, 373 (1993).

\end{thebibliography}
\end{document}